\documentclass[aps,prb,twocolumn,superscriptaddress]{revtex4-2}
\usepackage{amsmath,amssymb,graphicx,bm}
\usepackage{hyperref}

\usepackage{booktabs}
\usepackage{threeparttable}
\usepackage{rotating}
\begin{document}
\title{
Unified Transport and Susceptibility Analysis of a Thin BSCCO Film: From Local Pairing to Global Phase Coherence \\[0.5em]
}
\author{Santu Prasad Jana}
\thanks{Corresponding author: santup@iisc.ac.in}
\affiliation{Centre for Nanoscience and Engineering, Indian Institute of Science, Bengaluru, Karnataka 560012, India}
\author{Bismaya Ranjan Nayak}
\affiliation{Department of Physics, Indian Institute of Science, Bengaluru, Karnataka 560012, India}
\author{Sohini Guin}
\affiliation{Centre for Nanoscience and Engineering, Indian Institute of Science, Bengaluru, Karnataka 560012, India}
\author{Akshay Naik}
\affiliation{Centre for Nanoscience and Engineering, Indian Institute of Science, Bengaluru, Karnataka 560012, India}
\author{Dhavala Suri}
\affiliation{Centre for Nanoscience and Engineering, Indian Institute of Science, Bengaluru, Karnataka 560012, India}
\author{Arindam Ghosh}
\affiliation{Department of Physics, Indian Institute of Science, Bengaluru, Karnataka 560012, India}
\date{\today}

\begin{abstract}
The superconducting transition in thin high-$T_c$ films can be significantly broadened by disorder, spatial inhomogeneity, and phase fluctuations. Here, we study the transition in a 30-nm-thick sputter-grown Bi$_2$Sr$_2$CaCu$_2$O$_{8+\delta}$ (BSCCO) film through electrical-transport and ac-susceptibility measurements. The resistive transition is examined using a Gaussian distribution of local $T_c$ values and the Ambegaokar--Halperin model, describing spatial variations in superconductivity and thermally activated phase dynamics, respectively. To relate the transport and magnetic responses, we extend the Choy--Stoneham susceptibility model by introducing the superconducting fraction extracted from transport and a temperature-dependent connectivity factor representing the gradual development of Josephson coupling among superconducting regions. The results identify separate temperature regimes corresponding to the onset of local superconductivity, expansion of the superconducting fraction, development of magnetic screening, and establishment of global phase coherence. The transition therefore proceeds progressively, from locally superconducting regions to a connected, phase-coherent state. This analysis provides a common framework for relating spatial inhomogeneity, dissipative phase dynamics, and magnetic screening in inhomogeneous superconducting films.
\end{abstract}
\maketitle

\section{INTRODUCTION}
The relationship between disorder, spatial inhomogeneity, and superconductivity remains an important problem in condensed-matter physics \cite{McElroy,disorder}. In a homogeneous superconductor, the formation of Cooper pairs and the establishment of macroscopic phase coherence are generally associated with a well-defined superconducting transition \cite{NbN}, accompanied by the onset of zero resistance and diamagnetic screening \cite{Tinkham}. This picture can change considerably in low-dimensional and strongly correlated superconductors, where disorder, variations in carrier density, strain, and structural defects produce spatial variations in the superconducting properties \cite{STM,STM1,STM3}. Such effects are especially relevant in thin high-$T_c$ cuprate films because of their short coherence length \cite{BSCCO film}, strong anisotropy \cite{anisotropy,anisotropy1,anisotropy2}, quasi-two-dimensional electronic structure, and sensitivity to oxygen stoichiometry \cite{High Tc oxide}.

Under these conditions, the superconducting state may develop over an extended temperature interval rather than through a single sharp transition \cite{Review,Tinkham1,Zhang2023,Mua,Aytekin2024,Aichner,Yu2019}. Local superconducting regions can appear while the sample remains resistive, with global phase coherence emerging only at a lower temperature as coupling between these regions becomes sufficiently strong. Identifying how the system evolves from local superconductivity to a macroscopically coherent state is therefore essential for understanding broadened transitions in disordered cuprates.

Two related mechanisms are commonly invoked to explain this broadening. The first is spatial inhomogeneity, which produces a distribution of local superconducting transition temperatures across the sample \cite{Tc distribution,Tc distribution1,Tc distribution2}. Different regions then become superconducting at different temperatures, causing the superconducting fraction to increase continuously upon cooling. The second involves fluctuations of the superconducting phase. Thermally activated phase dynamics, vortex motion, and weak Josephson coupling can maintain a finite resistance even after the local pairing amplitude has developed \cite{AH,phase fluctuation,phase slip,phase slip2}. In this regime, the resistive transition reflects not only the formation of superconducting regions but also the gradual establishment of phase coherence between them.

Distinguishing these contributions using transport measurements alone is difficult because the measured resistance depends on both the spatial distribution of superconductivity and the connectivity of the superconducting paths. Magnetic susceptibility provides complementary information by probing the formation of screening currents and the collective electromagnetic response of the superconducting regions \cite{Kumar2013}. The onset of a local superconducting fraction, the formation of a continuous low-resistance path, and the development of an appreciable diamagnetic response need not occur at the same temperature. A combined analysis of transport and susceptibility can therefore resolve aspects of the transition that are not readily accessible from either measurement separately.

Bi$_2$Sr$_2$CaCu$_2$O$_{8+\delta}$ (BSCCO) is particularly suitable for examining these effects. Its layered crystal structure, strong electronic anisotropy \cite{anisotropy1}, and quasi-two-dimensional character enhance the role of spatial variations and phase fluctuations \cite{McElroy,superconducting fluctuations}. In thin films, these intrinsic properties can be further influenced by strain, finite thickness, structural disorder, and spatial variations in oxygen content. Scanning tunneling microscopy and spectroscopy measurements have directly revealed nanoscale variations in the electronic structure and superconducting gap of BSCCO \cite{STM,STM1,Cooper pairing,STM2}. Beyond its fundamental interest, BSCCO is relevant to high-$T_c$ superconducting devices, including SQUIDs and emerging photon- and kinetic-inductance-based detectors, whose performance depends sensitively on the uniformity and stability of the superconducting state. Although broadened transitions are frequently observed in such systems, their transport and magnetic responses are often treated independently. Consequently, the connection among the local onset of superconductivity, inter-region phase coherence, and macroscopic magnetic screening remains incompletely understood. 

In this work, we investigate the superconducting transition of a 30-nm-thick BSCCO film using electrical-transport and ac-susceptibility measurements. We analyze the resistive transition using a Gaussian distribution of local $T_c$ values and the Ambegaokar--Halperin framework. These two descriptions emphasize spatial variations in the local transition temperature and thermally activated phase dynamics, respectively. We then extend the Choy--Stoneham susceptibility model by incorporating the superconducting fraction $C_s(T)$ obtained from the transport analysis and a temperature-dependent connectivity factor $J(T)$. The product $C_s(T)J(T)$ defines the effective superconducting fraction that contributes to the measured magnetic response. The combined analysis identifies characteristic temperature scales associated with the local onset of superconductivity, growth of the superconducting fraction, magnetic screening, and global phase coherence. The Gaussian-$T_c$ analysis is further applied to reported Bi-2212 systems, while the transport--susceptibility framework is examined using NbTiN and nanoporous NbN as reference systems, placing the present results in the broader context of inhomogeneous superconductivity.

\section{EXPERIMENTAL DETAILS}
\begin{figure}
	\centering
 	\includegraphics[width=3.4in]{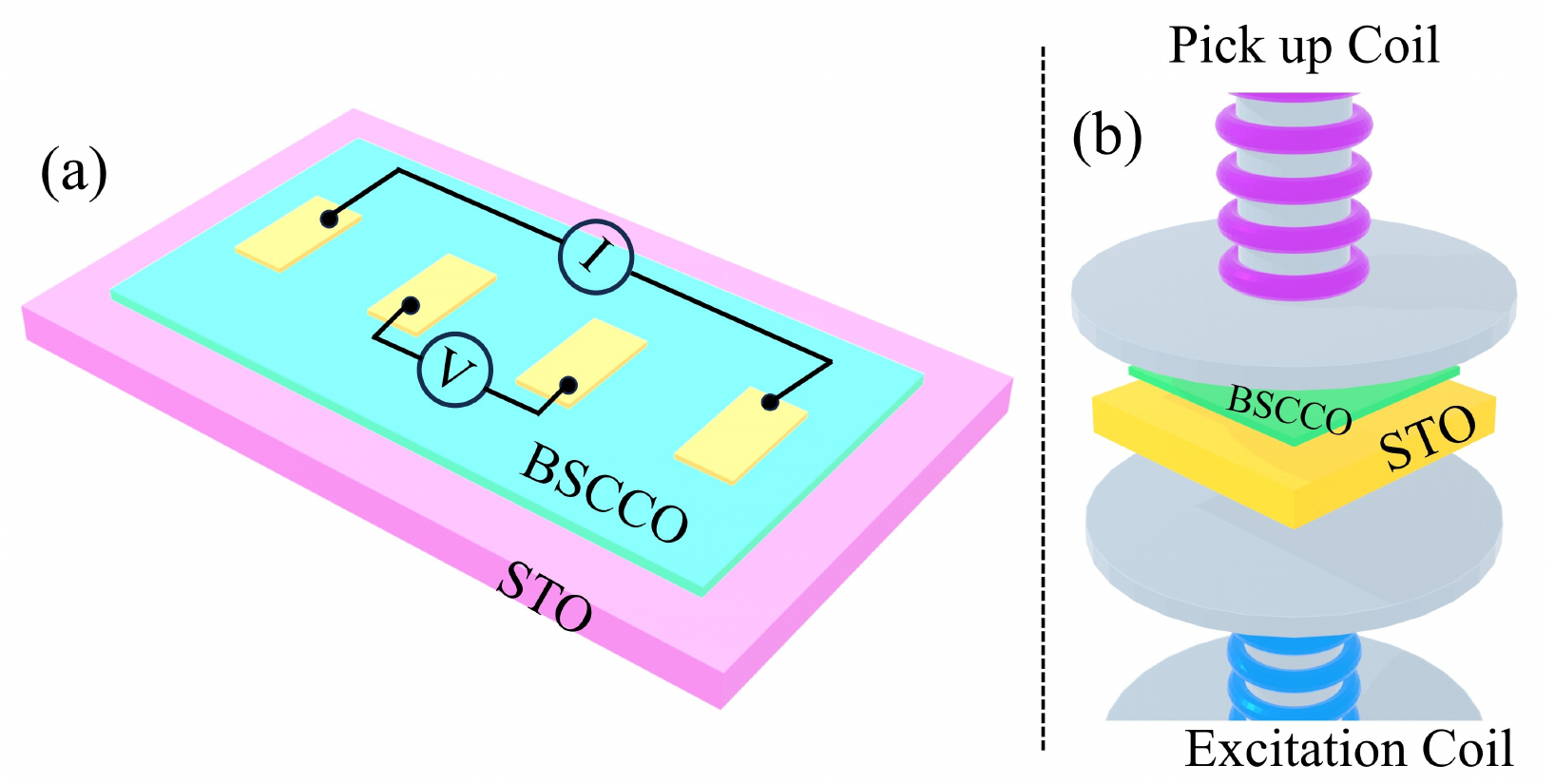}
	\caption{Experimental configurations for (a) four-probe resistance--temperature (\(R\)-\(T\)) measurements and (b) AC magnetic susceptibility measurements of the BSCCO thin film on an STO substrate.}
	\label{fig:fig9}
\end{figure}

BSCCO thin films were deposited on $5\times5$~mm$^2$ STO (100) substrates by RF sputtering at room temperature. Deposition was performed at 100~W under an Ar flow of 40~sccm, with base and working pressures of approximately $10^{-7}$ and $10^{-3}$~mbar, respectively, yielding a growth rate of $8$--$10$~nm/min. The films were subsequently annealed in an O$_2$ atmosphere at $850\,^{\circ}\mathrm{C}$ for 1~h. Temperature-dependent four-probe DC resistance measurements were performed in a cryocooler using a Keithley 6221 as the DC current source and a Keithley 2181A nanovoltmeter for voltage measurements, following the measurement configuration shown in Fig. \ref{fig:fig9}(a).

\par The ac susceptibility of the same film was measured using the mutual-inductance technique illustrated in Fig.~\ref{fig:fig9}(b). An excitation current of amplitude $I_0$ and frequency $\nu$ produces a pickup-voltage amplitude $V(\nu)=2\pi\nu I_0M$, where $M$ is the mutual inductance between the excitation and pickup coils and contains the magnetic response of the sample~\cite{Kumar2013}. The relation between the pickup voltage and the ac susceptibility also depends on geometrical factors, including the filling factor and pickup-coil coupling. Because the overall proportionality factor could not be determined accurately for the present experimental configuration, the measured in-phase voltage is reported in arbitrary susceptibility units.

Accordingly, the overall proportionality factor is absorbed into the susceptibility scale, and the analysis focuses on its temperature dependence and relative variation rather than its absolute magnitude. This treatment does not affect the determination of the superconducting transition or the subsequent analysis. Before measuring the BSCCO film, the reliability of the setup was independently verified using an approximately 25-nm-thick sputter-grown NbTiN reference film. Its temperature-dependent resistance and susceptibility, together with the corresponding analyses, are presented in the Supplemental Material~\cite{Suppl}.

\section{THEORETICAL Model}
\subsection{Gaussian $T_c$ Distribution Model}
\begin{figure*}
    \centering
    \includegraphics[width=6.6in]{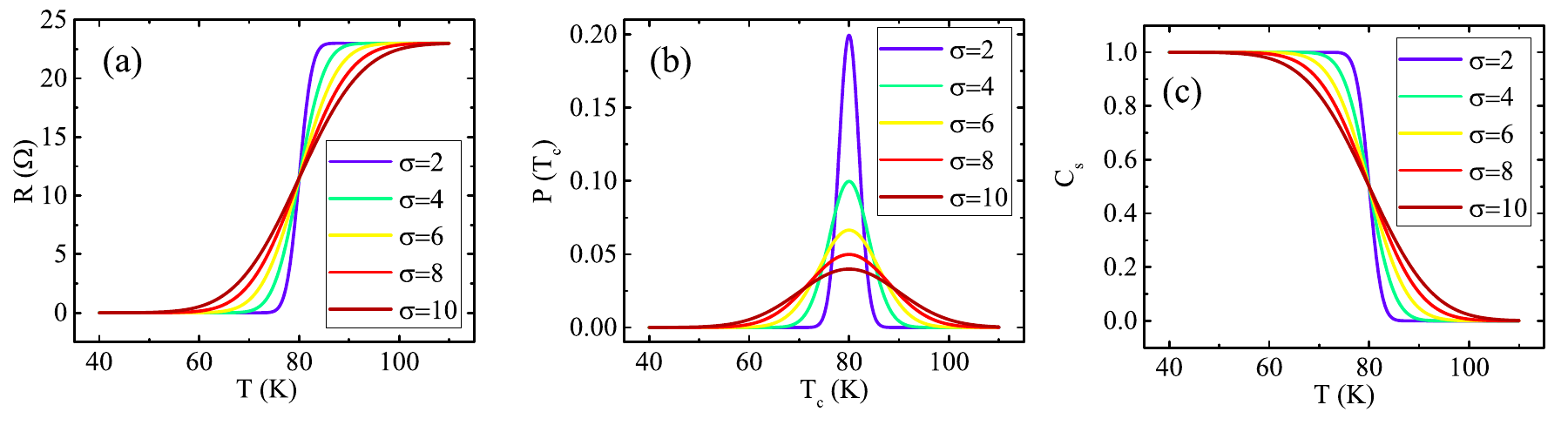}
    \caption{Superconducting transition simulated using a Gaussian distribution of local transition temperatures. (a) Resistance $R(T)$ calculated for different standard deviations $\sigma$. (b) Corresponding probability distributions $P(T_c)$. (c) Superconducting fraction $C_s(T)$. Increasing $\sigma$ broadens the resistive transition and produces a more gradual development of the superconducting fraction, while the mean transition temperature $T_{c0}$ remains fixed.}
    \label{fig:fig1}
\end{figure*}

The superconducting transition of a thin BSCCO film can be broadened by spatial variations in its superconducting properties. BSCCO is highly anisotropic and is characterized by short zero-temperature coherence lengths, typically $\xi_{ab}(0)\sim2$~nm within the CuO$_2$ planes and $\xi_c(0)\sim0.1$--$0.3$~nm along the crystallographic $c$ axis \cite{anisotropy1}. The superconducting state is therefore sensitive to local variations occurring on nanometer length scales. In sputter-grown thin films, spatial variations in oxygen content \cite{Oxygen Dopants}, carrier concentration \cite{doping}, strain, and defect density can produce superconducting regions characterized by different local transition temperatures $T_c(\mathbf{r})$.

\par Upon cooling, regions with higher local $T_c$ become superconducting before those with lower $T_c$. From a transport perspective, the measured resistance reflects an effective-medium response arising from a large ensemble of such locally superconducting and normal regions. Global zero resistance is achieved only when a continuous superconducting path percolates across the film \cite{percolative}. The superconducting fraction consequently develops over a finite temperature interval rather than at a single temperature, produces a smooth and extended resistive transition.

\par According to the central limit theorem, the cumulative effect of many independent and weakly correlated random contributions can, in general, lead to an approximately normal (Gaussian) distribution \cite{central limit}. In the present context, the local superconducting transition temperature is influenced by multiple microscopic factors, including nanoscale doping fluctuations \cite{Oxygen Dopants,doping}, structural disorder \cite{disorder}, strain fields, and interface-related effects. While these perturbations may not be strictly independent and can involve correlations or percolative effects, they introduce random variations in the superconducting properties, such that the resulting distribution of local $T_c$ values across the film can be reasonably approximated by a Gaussian form :
\begin{equation}
P(T_c) = \frac{1}{\sqrt{2\pi}\sigma}
\exp\!\left[-\frac{(T_c - T_{c0})^2}{2\sigma^2}\right],
\label{equ:equ1}
\end{equation}
where $T_{c0}$ represents the mean superconducting transition temperature of the film and $\sigma^2 = \left\langle \left( T_c - T_{c0} \right)^2 \right\rangle$ is the standard deviation of the distribution, represents the rms spatial variation of the local superconducting transition temperature across the film.
 
\par  At a given temperature $T$, regions with local $T_c(\mathbf{r}) > T$ are superconducting and contribute negligibly to the resistance, whereas regions with $T_c(\mathbf{r}) < T$ remain in the normal state and contribute their normal-state resistance. The total measured resistance can therefore be expressed as a statistical average over the fraction of regions that remain normal:
\begin{equation}
R(T) = R_n \int_{-\infty}^{T} P(T_c)\, dT_c,
\label{equ:equ2}
\end{equation}
where $R_n$ is the normal-state resistance of the film. Substituting the Gaussian form of $P(T_c)$ into this expression yields
\begin{equation}
R(T) = \frac{R_n}{2}
\left[
1 + \operatorname{erf}\!\left(\frac{T - T_{c0}}{\sqrt{2}\sigma}\right)
\right],
\label{equ:equ3}
\end{equation}
resulting in an error-function-like \cite{error function} temperature dependence of the resistive transition.

The superconducting fraction is obtained by integrating over regions with $T_c>T$:
\begin{equation}
C_s(T)=\int_{T}^{\infty}P(T_c)\,dT_c,
\label{equ:equ4}
\end{equation}
which yields
\begin{equation}
C_s(T)=\frac{1}{2}
\left[
1-\operatorname{erf}\!\left(
\frac{T-T_{c0}}{\sqrt{2}\sigma}
\right)
\right].
\label{equ:equ5}
\end{equation}
Within the assumptions of the model, Eqs.~(\ref{equ:equ3}) and (\ref{equ:equ5}) are related by $C_s(T)=1-R(T)/R_n$. Thus, $C_s(T)$ evolves from approximately zero above the transition to unity well below it.

Figure~\ref{fig:fig1} illustrates the effect of varying $\sigma$ while keeping $T_{c0}$ and $R_n$ fixed. As shown in Fig.~\ref{fig:fig1}(a), increasing $\sigma$ from 2 to 10~K broadens the resistive transition, transforming a sharp, nearly step-like decrease in resistance into a smooth and extended crossover without shifting the transition midpoint. The corresponding distributions in Fig.~\ref{fig:fig1}(b) become broader and decrease in peak amplitude, indicating a larger spread of local transition temperatures and enhanced spatial inhomogeneity in the superconducting properties. Fig.~\ref{fig:fig1}(c) shows that the same increase in $\sigma$ produces a more gradual evolution of $C_s(T)$.

 These simulations demonstrate that a Gaussian distribution of local $T_{c}$ naturally produces both a broadened resistive transition and a gradual development of the superconducting fraction. The quantity $c_s(T)$ extracted from this model plays a central role in the subsequent analysis, where it is used not only to describe the transport behavior but also to establish a direct connection between electrical transport, Josephson connectivity, and magnetic susceptibility in the thin BSCCO film.

\subsection{Ambegaokar-Halperin Model}
A phase slip is a process in which the superconducting phase difference changes by $2\pi$, accompanied by a transient local suppression of the order-parameter amplitude \cite{Tinkham,AH}. In a quasi-two-dimensional superconducting film, an equivalent phase change can occur when a vortex crosses the sample transverse to the applied current \cite{Vortex pinning,phase slip,phase slip1}. The vortex crossing changes the phase difference across the sample by $2\pi$ and generates a voltage pulse according to the Josephson relation $V=(\hbar/2e)\,d\phi/dt$, where $\hbar$ is the reduced Plank constant and $e$ is the electronic charge. If $\Gamma$ is the thermally activated phase-slip (TAPS) or vortex crossing rate \cite{phase slip,phase slip1}, then the average voltage becomes $\langle V \rangle = (\hbar/2e)(2\pi \Gamma) = \Phi_0 \Gamma$, where $\Phi_0 = h/2e$ is the superconducting flux quantum. This produces a finite measurable resistance even below the mean-field $T_c$, despite the presence of a superconducting gap and Cooper pairing.

\par The Ambegaokar--Halperin (AH) model describes thermally activated phase-slip events, whose repeated occurrence leads to phase diffusion in an overdamped Josephson junction (JJ) \cite{Tinkham,AH}. Although a BSCCO film is not a single Josephson junction, its strongly anisotropic and layered structure allows it to be viewed as an effective network of Josephson-coupled regions arising from the intrinsic stacking of CuO$_2$ planes and additional disorder-induced weak links \cite{Intrinsic JJ,Intrinsic JJ1}. Thermal activation across these links can generate dissipative phase dynamics, while vortex motion provides a related mechanism for phase relaxation in the extended film. Structural disorder, defects, and grain boundaries can act as pinning centers that impede vortex motion. A vortex crossing the film must overcome a free-energy barrier $\Delta U$, leading near the superconducting transition to thermally activated vortex hopping with a rate $\Gamma\propto\exp[-\Delta U/(k_BT)]$.

\par Within the AH framework, this dissipative behavior is represented phenomenologically by thermally activated phase slips across an effective Josephson weak link. At zero applied bias, the corresponding washboard-potential barrier is $\Delta U\approx2E_J$~\cite{AH}, where $E_J(T)=\hbar I_c(T)/(2e)$ is the Josephson coupling energy and $I_c(T)$ is the effective temperature-dependent critical current of the weak-link network. Close to the transition, we approximate the critical current as $I_c(T)=I_c(0)(1-T/T_c)$ for $T<T_c$, where $I_c(0)$ is the extrapolated zero-temperature effective critical current. The resulting values of $E_J$ and $I_c$ provide effective measures of the phase stiffness of the weak-link network rather than a complete microscopic description of the film. Such effective AH-type approaches have also been used to describe vortex dynamics and thermally activated phase-slip dissipation in polycrystalline high-$T_c$ cuprates \cite{phase slip2,phase slip,dissipation}, superconducting islands on graphene \cite{phase slip1}.

\par In the zero-bias linear-response limit, the AH resistance is given by
\begin{equation}
R(T)=\frac{R_N}
{I_0^2\!\left(\frac{E_J(T)}{k_BT}\right)},
\label{equ:equ6}
\end{equation}
where $R_N$ is the normal-state resistance and $I_0(x)$ is the zeroth-order modified Bessel function \cite{AH}. The dimensionless ratio $E_J/(k_BT)$ determines the competition between Josephson coupling and thermal fluctuations. When $E_J\gg k_BT$, strong phase locking suppresses the resistance. When $E_J\lesssim k_BT$, thermally activated phase diffusion produces appreciable dissipation, while in the limit $E_J\ll k_BT$, the resistance approaches $R_N$. The model therefore describes a continuous crossover from a thermally fluctuating resistive regime to a phase-coherent state upon cooling.

\subsection{Effective Superconducting Fraction and Magnetic Susceptibility}
The magnetic response of a spatially inhomogeneous superconductor is governed by both the formation of local superconducting regions and the establishment of phase coherence between them. In systems exhibiting a distribution of local superconducting properties, superconductivity develops progressively upon cooling, resulting in a broadened diamagnetic transition rather than an abrupt onset of the Meissner state. To describe such behavior, an effective-medium approach can be employed in which the macroscopic magnetic response is determined by the temperature-dependent fraction of superconducting regions and their ability to screen magnetic flux. According to the effective-medium formulation developed by Choy and Stoneham \cite{Choy and Stoneham}, the magnetic susceptibility is expressed as

\begin{equation}
\chi(T)=
\frac{1}{4\pi}
\left[
\frac{1-c_{\rm s}(T)Z_{\lambda}(T)}
{1+\frac{1}{2}c_{\rm s}(T)Z_{\lambda}(T)}
-1
\right],
\label{eq:equ7}
\end{equation}

where $c_{\rm s}(T)$ represents the superconducting fraction and $Z_{\lambda}(T)$ is a magnetic shielding factor that accounts for finite magnetic-field penetration into superconducting regions. 

\par In the present work, we extend this formulation by introducing an effective superconducting fraction
\begin{equation}
c_{\rm eff}(T)=c_s(T)J(T),
\label{equ:equ8}
\end{equation}
where $c_s(T)$ obtained from the Gaussian distribution of local transition temperatures and $J(T)$ is a connectivity factor describing the progressive establishment of long-range phase coherence between superconducting regions. Substituting $c_{\rm eff}(T)$ for $c_s(T)$ yields
\begin{equation}
\chi(T)=
\frac{1}{4\pi}
\left[
\frac{1-c_{\rm eff}(T)Z_{\lambda}(T)}
{1+\frac{1}{2}c_{\rm eff}(T)Z_{\lambda}(T)}
-1
\right].
\label{equ:equ9}
\end{equation}
This modification accounts for the fact that local superconductivity may emerge at temperatures significantly above the onset of bulk diamagnetic screening, which requires coherent coupling between superconducting regions.

\par The magnetic shielding factor $Z_{\lambda}(T)$ is obtained from the solution of the London equation for a superconducting region of characteristic size $a$:
\begin{equation}
Z_{\lambda}(T)=
1-\frac{3}{x}\coth(x)+\frac{3}{x^{2}},
\label{eq:Zlambda}
\end{equation}
where $x=a/\lambda(T)$. The temperature-dependent penetration depth is described using the Ginzburg--Landau form $\lambda(T)=\lambda_0/\sqrt{1-T/T_{c0}}$ for $T<T_{c0}$, where $\lambda_0$ is the penetration-depth scale and $T_{c0}$ is the mean transition temperature. The factor $Z_{\lambda}(T)$ accounts for the reduction of the ideal Meissner response due to magnetic-field penetration into a superconducting region. In the limit $a\gg\lambda(T)$, the magnetic field is largely expelled and $Z_{\lambda}(T)\rightarrow1$. Conversely, when $a\lesssim\lambda(T)$, substantial field penetration reduces the diamagnetic response.

Although originally developed for granular superconductors, the effective-medium framework provides a useful phenomenological description for ultrathin superconducting films exhibiting spatially non-uniform superconductivity. In such systems, local variations in disorder, oxygen content, strain, or carrier concentration can produce regions with different superconducting characteristics. The quantity $c_s(T)$ describes the progressive emergence of these locally superconducting regions, while the connectivity factor $J(T)$ captures the subsequent development of phase coherence between them. The resulting susceptibility therefore reflects the combined effects of superconducting fraction, phase connectivity, and magnetic screening, providing a direct link between transport and magnetic measurements within a unified framework.

\section{RESULTS AND ANALYSIS}
\subsection{Gaussian $T_c$ Analysis}
\begin{figure*}
\centering
\includegraphics[width=6.2in]{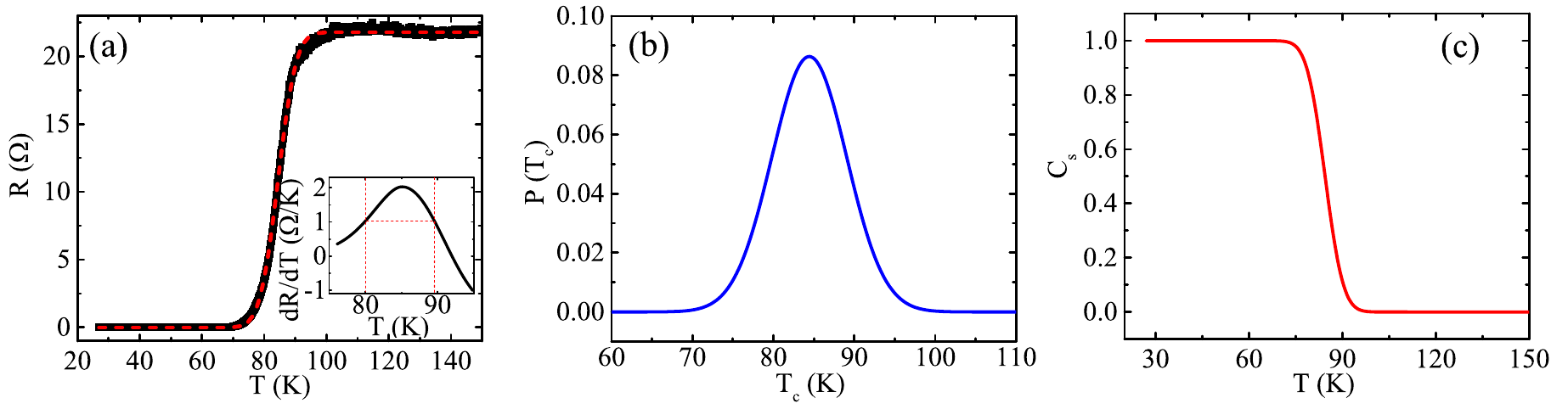}
\caption{Gaussian-distribution analysis of the superconducting transition. (a) Experimental resistance $R(T)$ (black circles) and fit obtained using the cumulative Gaussian model (red dashed line). The inset shows the temperature derivative $dR/dT$; the horizontal and vertical dashed lines indicate the half-maximum level and corresponding temperatures used to determine the full width at half maximum (FWHM) of the derivative peak. (b) Probability distribution $P(T_c)$ of local superconducting transition temperatures obtained from the fitted parameters. (c) Corresponding superconducting fraction $C_s(T)$ calculated from the fitted $T_c$ distribution.}
\label{fig:fig2}
\end{figure*}

Figure~\ref{fig:fig2}(a) presents the experimental $R(T)$ data together with the fit obtained using Eq.~(\ref{equ:equ3}), which describes the transition in terms of a Gaussian distribution of local critical temperatures. The normal-state resistance $R_n$, mean transition temperature $T_{c0}$, and distribution width $\sigma$ were treated as fitting parameters. The model reproduces the measured transition closely, yielding $\chi_{\mathrm{red}}^2=1.034$ and $R^2=0.99972$. The extracted parameters are $T_{c0}=84.44\pm0.01$~K, $\sigma=4.62\pm0.01$~K, and $R_n=21.76~\Omega$. The Gaussian distribution $P(T_c)$ calculated from these parameters is shown in Fig.~\ref{fig:fig2}(b). Its finite width indicates a distribution of local transition temperatures that accounts for the broadening of the measured $R(T)$ transition within the present model.

\par The corresponding superconducting fraction $C_s(T)$ is shown in  Fig.~\ref{fig:fig2}(c). It increases gradually from nearly zero to unity over approximately 92--70~K, reflecting the progressive development of superconductivity across regions with different local $T_c$ values. By definition, $C_s(T)=0.5$ at $T=T_{c0}$, consistent with the center of the fitted distribution.

\par For a Gaussian $T_c$ distribution, the 90--10\% resistive-transition width $\Delta T_{90\text{--}10}$ is related to $\sigma$ by $\Delta T_{90\text{--}10}\approx2.56\sigma$, as derived in Appendix~A. The fitted value $\sigma=4.62$~K therefore gives $\Delta T_{90\text{--}10}\approx11.8$~K, in agreement with the width determined directly from the experimental $R(T)$ data. This agreement demonstrates that $\sigma$ provides a physically interpretable measure of the transition broadening within the Gaussian-distribution framework.

\par Figure~\ref{fig:fig2}(a) inset shows the numerically smoothed derivative $dR/dT$. The derivative exhibits a single, approximately symmetric peak centered near $T_{c0}$. This behavior is consistent with the Gaussian model because differentiating the error-function form of $R(T)$ yields a Gaussian temperature dependence. The peak width provides an additional consistency check on the fitted distribution. Using $\mathrm{FWHM}=2.3548\sigma$, as derived in Appendix~A, the measured $\mathrm{FWHM}=9.85$~K corresponds to $\sigma\approx4.2$~K, reasonably close to the fitted value of 4.62~K. The small difference may arise from numerical smoothing, experimental noise, and deviations of the measured transition from an ideal Gaussian form. Overall, the agreement between the $R(T)$ fit and the derivative analysis supports the internal consistency of the Gaussian $T_c$-distribution model.

\par
\begin{figure}
    \centering
    \includegraphics[width=3.2in]{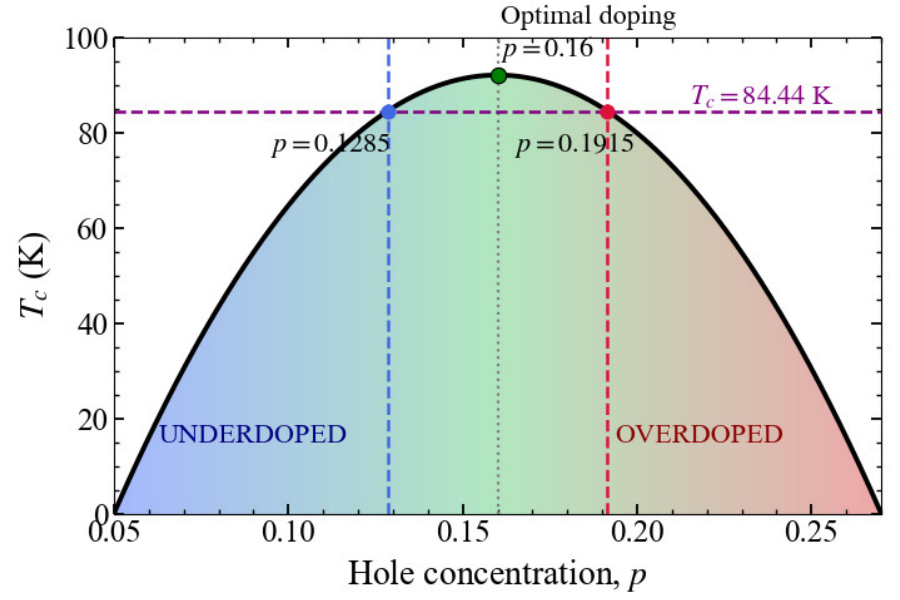}
    \caption{Empirical $T_c$--$p$ superconducting dome for Bi$_2$Sr$_2$CaCu$_2$O$_{8+\delta}$ calculated using the Presland relation. The measured value $T_{c0}=84.44$~K gives two possible hole concentrations, $p=0.1285$ (underdoped) and $p=0.1915$ (overdoped), located symmetrically about the optimal value $p=0.16$.}
    \label{fig:fig8}
\end{figure}

 To gain further insight into the possible origin of the transition broadening, the extracted mean transition temperature was examined using the empirical Presland relation, $T_c/T_c^{\max}=1-82.6(p-0.16)^2$, which relates the superconducting transition temperature to the hole concentration $p$ per planar Cu atom in cuprate superconductors \cite{Presland1991}. Differentiating this relation gives $dT_c/dp=-165.2T_c^{\max}(p-0.16)$. Using $T_{c0}=84.44$~K and $T_c^{\max}\approx92$~K for BSCCO yields an effective deviation from optimal doping of $|p-0.16|\approx0.0315$. Because the Presland relation is symmetric about optimal doping, two possible solutions are obtained: $p=0.1285$ on the underdoped side and $p=0.1915$ on the overdoped side, as illustrated in Fig.~\ref{fig:fig8}. If the reduction in $T_c$ is primarily associated with oxygen deficiency in the sputter-grown film, the underdoped solution, $p\approx0.129$, is the more plausible of the two.

\par Assuming that $T_c$ varies smoothly with $p$, small spatial variations in the hole concentration can be related to variations in the local transition temperature through linear error propagation, $\sigma_{T_c}=|dT_c/dp|\,\sigma_p$. At $p\approx0.1285$, the fitted width $\sigma_{T_c}=4.62$~K corresponds to $\sigma_p\approx0.010$. This value should be interpreted as an effective doping-equivalent measure of the transition inhomogeneity rather than as the exact microscopic variation in hole concentration. In a 30-nm-thick BSCCO film, spatial variations in $T_c$ may also arise from structural disorder, strain, oxygen nonstoichiometry, interface effects, and finite-thickness effects. The estimated value of $\sigma_p$ is nevertheless compatible with the nanoscale electronic inhomogeneity reported in scanning tunneling microscopy, spectroscopy, and related measurements of BSCCO \cite{STM,Tallon2001,McElroy}, supporting a description of the broadened transition in terms of spatially varying local superconducting properties.

\par To obtain an order-of-magnitude estimate of the spatial scale associated with the inferred doping inhomogeneity, we consider disorder correlated over an in-plane length $l_d$. A superconducting coherence area $A_\xi\sim\xi_{ab}^2$ then contains approximately $N_\xi\sim A_\xi/l_d^2$ statistically independent disorder regions. If each region is characterized by a local doping variation $\delta p_0$, spatial averaging over the coherence area reduces the effective variation to $\sigma_p\sim\delta p_0/\sqrt{N_\xi}$. Substitution of $N_\xi$ gives $\sigma_p\sim\delta p_0l_d/\xi_{ab}$, and hence
$l_d\sim\xi_{ab}\sigma_p/\delta p_0$.

\par Taking $\xi_{ab}\sim1.5$~nm, $\sigma_p\approx0.010$, and a representative local doping variation $\delta p_0\approx0.02$--$0.03$ \cite{Tallon2001,McElroy} gives $l_d\sim0.5$--$0.75$~nm. This estimate corresponds to a length scale of only a few in-plane lattice spacings and is compatible with disorder associated with oxygen dopants and structural variations in the BiO layers \cite{dopant}. Since the calculation neglects geometrical factors, spatial correlations, and other sources of $T_c$ suppression, $l_d$ should be regarded only as an effective order-of-magnitude scale rather than a direct measurement of the microscopic disorder-correlation length.

\subsection{AH Analysis}
\begin{figure*}
    \centering
    \includegraphics[width=6.2in]{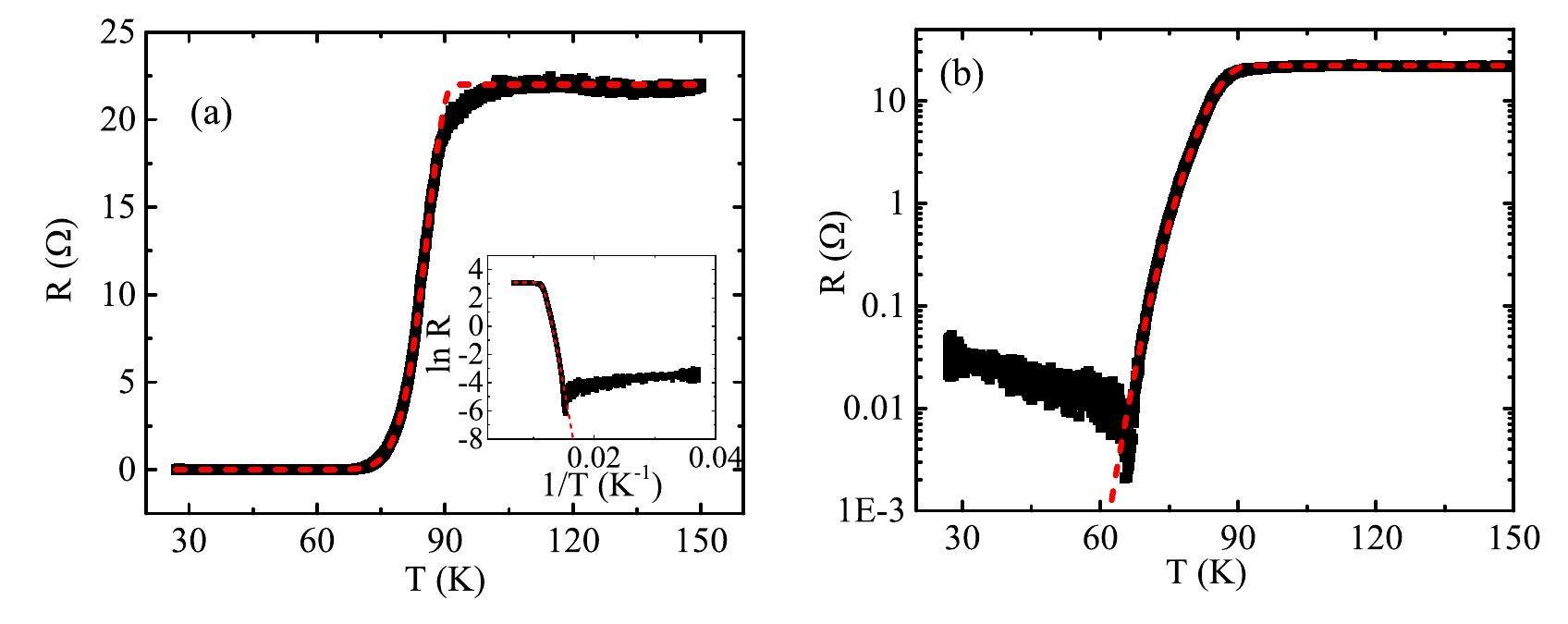}
    \caption{Comparison of the measured resistance with the Ambegaokar--Halperin model. (a) Resistance $R(T)$ on a linear scale, showing the experimental data (black symbols) and AH fit (red dashed line). The inset shows an Arrhenius representation of $\ln R$ as a function of $1/T$; the linear fit is used to estimate an effective activation energy. (b) Same data and AH fit on a semilogarithmic scale, emphasizing the low-resistance region of the transition.}
    \label{fig:fig3}
\end{figure*}

To examine the role of thermally activated phase dynamics in the broadened superconducting transition, the resistance data were analyzed using the Ambegaokar--Halperin (AH) model introduced in the preceding section. Figure~\ref{fig:fig3}(a) compares the experimental $R(T)$ data with the AH fit, while Fig.~\ref{fig:fig3}(b) presents the same results on a semilogarithmic scale to emphasize the low-resistance regime. The normal-state resistance $R_N$, transition temperature $T_c$, and extrapolated zero-temperature critical current $I_{c0}$ were treated as fitting parameters. The model captures the principal features of the transition, yielding $T_c=92$~K, $R_N=22~\Omega$, and $I_{c0}=5.5\times10^{-5}$~A. At lower temperatures, the model predicts a continued decrease in resistance, whereas the measured signal approaches a finite floor. This deviation may arise from the voltage-resolution limit of the measurement setup and/or additional residual dissipation not included in the AH model.

\par  Here, $I_{c0}$ should not be identified with the intrinsic depairing current of the superconducting condensate \cite{dissipation}; rather, it defines an effective current scale for phase locking across the weak-link network and is governed primarily by the least strongly coupled regions controlling global coherence. Within the same effective description and the assumed temperature dependence of $I_c(T)$, the fitted value $T_c=92$~K represents the temperature at which the Josephson coupling energy extrapolates to zero. It is therefore interpreted as an effective mean-field or local-pairing scale rather than the temperature at which global zero resistance is established. This value is close to the maximum transition temperature reported for optimally doped BSCCO-2212 \cite{Presland1991}. In comparison, the Gaussian-distribution analysis yields a mean local transition temperature $T_{c0}=84.44$~K, corresponding to the midpoint of the broadened resistive transition. The separation between these temperature scales is consistent with the presence of local superconducting correlations above the temperature at which macroscopic phase coherence develops.

\par The fitted critical current follows $I_c(T)=I_{c0}(1-T/T_c)$, and the corresponding Josephson coupling energy is $E_J(T)=\hbar I_c(T)/(2e)$. At 90~K, this gives $I_c\approx1.20\times10^{-6}$~A and $E_J\approx3.94\times10^{-22}$~J. Since $k_BT\approx1.24\times10^{-21}$~J at this temperature, the ratio $E_J/(k_BT)\approx0.32$ indicates that thermal fluctuations dominate over the effective Josephson coupling close to $T_c$. Upon cooling, this ratio increases and reaches unity at $T^\ast\approx86.0$~K, close to the mean transition temperature $T_{c0}=84.44$~K obtained from the Gaussian analysis. This correspondence suggests that the central part of the resistive transition occurs in the temperature range where the effective Josephson coupling becomes comparable to the thermal energy. Below $T^\ast$, the increasing coupling progressively suppresses thermally activated phase fluctuations and promotes phase coherence between superconducting regions.

\par The activated behavior was further assessed from the Arrhenius plot of $\ln R$ versus $1/T$ shown in the inset of Fig.~\ref{fig:fig3}(a). Over the approximately linear fitting interval, the resistance was represented as $R\propto\exp[-\Delta U/(k_BT)]$, giving an effective activation energy $\Delta U=2.71\times10^{-21}$~J. This corresponds to $\Delta U/(k_BT_c)\approx2.1$, indicating that thermal activation remains appreciable in the transition region. In the zero-current limit of the AH model, the washboard-potential barrier is $\Delta U\approx2E_J$. Evaluating $2E_J$ over the central part of the Arrhenius fitting interval, $T=85$--$86$~K, gives approximately $(2.36$--$2.76)\times10^{-21}$~J, comparable to the Arrhenius value. This agreement provides a useful consistency check between the two analyses and supports thermally activated phase dynamics as an important contribution to the measured dissipation. The Arrhenius barrier should nevertheless be regarded as an effective value because $E_J$ varies with temperature and the AH resistance contains a temperature-dependent prefactor.

\subsection{Magnetic susceptibility analysis}

\begin{figure*}
    \centering
    \includegraphics[width=6.2in]{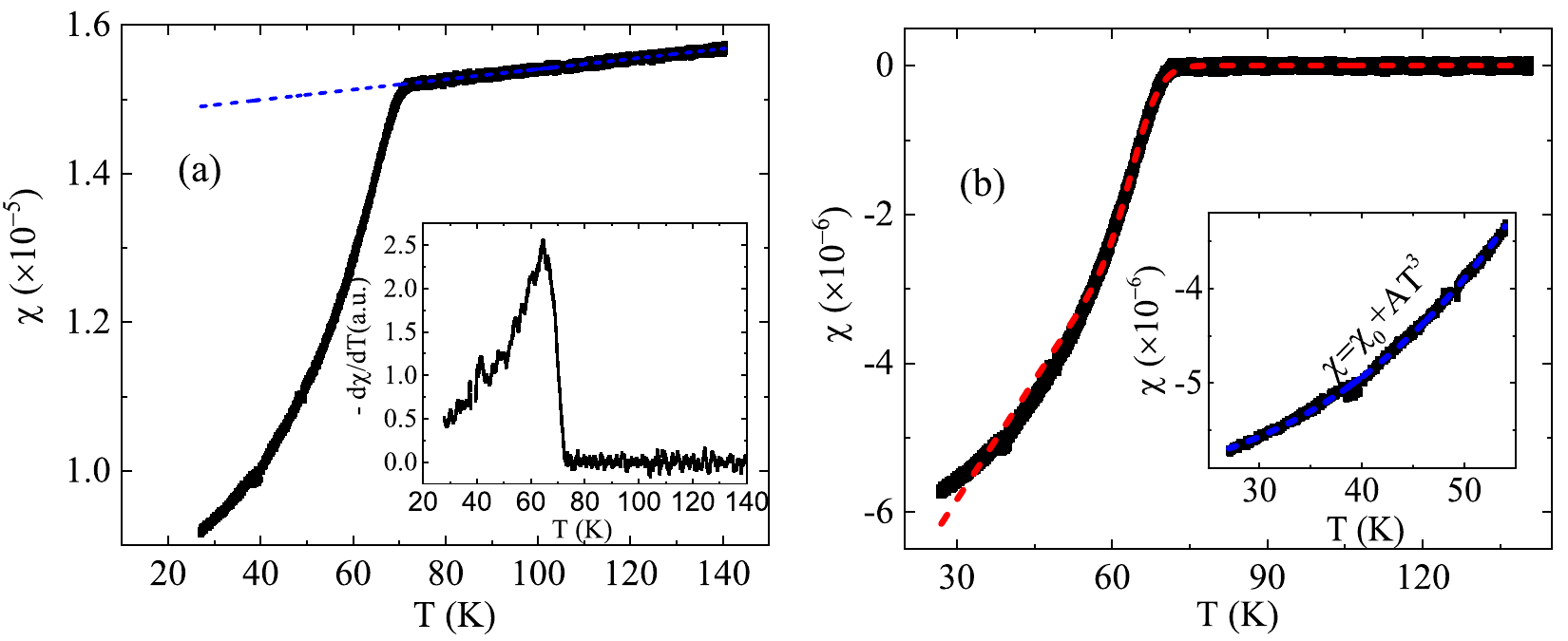}
    \caption{Magnetic response of the 30-nm-thick BSCCO film. (a) Raw susceptibility data (black symbols) and fitted normal-state background (blue dashed line). The inset shows the temperature derivative $-d\chi/dT$ of the background-subtracted susceptibility. (b) Background-subtracted susceptibility together with the fit obtained using the modified Choy--Stoneham model (red dashed line). The inset shows the low-temperature susceptibility and its empirical $T^3$ fit.}
    \label{fig:fig4}
\end{figure*}

Figure~\ref{fig:fig4}(a) shows the temperature dependence of the magnetic susceptibility measured with the excitation field applied perpendicular to the 30-nm-thick BSCCO film. The raw signal exhibits a gradual diamagnetic transition over a broad temperature interval, consistent with an inhomogeneous and progressively connected superconducting state. To isolate the superconducting contribution, the normal-state background was fitted using $\chi_{\mathrm{bg}}(T)=\chi_0+\alpha T$, yielding $\chi_0=1.472\times10^{-5}$~a.u. and $\alpha=6.884\times10^{-9}$~a.u.\,K$^{-1}$. The fitted background is represented by the blue dashed line in Fig.~\ref{fig:fig4}(a). The departure of the measured signal from this background marks the onset of detectable diamagnetic screening.

\par Figure~\ref{fig:fig4}(b) presents the background-subtracted susceptibility together with the modified Choy--Stoneham fit given by Eq.~(\ref{equ:equ9}). In this model, the superconducting fraction $C_s(T)$ of the original formulation \cite{Choy and Stoneham} is replaced by an effective fraction $C_{\mathrm{eff}}(T)=C_s(T)J(T)$. Here, $C_s(T)$ is obtained from the Gaussian analysis of the measured $R(T)$ transition [ Fig.~\ref{fig:fig2}(c)], whereas $J(T)=\frac{1}{2}\{1-\operatorname{erf}[(T-T_J)/\Delta_J]\}$ describes the gradual development of connectivity between superconducting regions. The product $C_s(T)J(T)$ therefore represents the effective superconducting fraction that contributes to the measured magnetic response and accounts for the delayed development of appreciable diamagnetic screening relative to the transport transition.

\par The fit gives a connectivity midpoint $T_J=65.32$~K and a crossover width $\Delta_J=6.08$~K. These values indicate that effective magnetic connectivity develops well below the mean transition temperature $T_{c0}=84.44$~K obtained from the transport analysis. The inset of Fig.~\ref{fig:fig4}(a) shows $-d\chi/dT$, which exhibits a pronounced peak near 65~K. This peak identifies the temperature at which the diamagnetic response changes most rapidly and agrees closely with the fitted value of $T_J$. Its asymmetric shape reflects the combined temperature dependences of the superconducting fraction, inter-region connectivity, and magnetic shielding. The agreement between the derivative peak and $T_J$ supports the consistency of the phenomenological connectivity description.

\par The shielding factor depends on the ratio $x(T)=a/\lambda(T)$, where $a$ is an effective characteristic length and $\lambda(T)$ is the penetration depth. With $\lambda(T)=\lambda_0/\sqrt{1-T/T_{c0}}$, this ratio can be written as $x(T)=x_0\sqrt{1-T/T_{c0}}$, where $x_0=a/\lambda_0$. The fit yields $x_0=0.0337$. Taking $\lambda_0\approx400$~nm, consistent with reported values for thin, disordered BSCCO films \cite{penetration depth,penetration depth1}, gives $a=x_0\lambda_0\approx13.5$~nm. Because the susceptibility depends primarily on $a/\lambda_0$, different combinations of $a$ and $\lambda_0$ can produce essentially the same calculated response. The inferred value of $a$ should therefore be regarded as an effective magnetic-screening length rather than a direct measurement of the physical grain or superconducting-domain size.

\par The susceptibility fit constrains the temperature evolution of magnetic screening more directly than the individual length scales $a$ and $\lambda_0$, which cannot be determined separately from the present data. By contrast, $T_J$ and $\Delta_J$ are governed mainly by the position and width of the diamagnetic crossover and are consequently less sensitive to this geometrical degeneracy. The effective-medium model captures the transition region but does not explicitly describe all aspects of the low-temperature penetration depth or superfluid response. Small systematic deviations from the fit at lower temperatures were therefore examined separately.

\par The inset of Fig.~\ref{fig:fig4}(b) shows the susceptibility over the range $27\leq T\leq54$~K. Within this interval, the data are well represented empirically by $\chi(T)=\chi_{\mathrm{LT},0}+AT^3$, with $\chi_{\mathrm{LT},0}=-6.03\times10^{-6}$~a.u., $A=1.71\times10^{-11}$~a.u.\,K$^{-3}$, and $R^2=0.9964$. In this temperature range, both $C_s(T)$ and $J(T)$ are nearly saturated within the model, so the remaining temperature dependence is not dominated by the growth of the superconducting fraction or connectivity. The observed $T^3$ behavior therefore characterizes the magnetic response of the established superconducting state over the measured interval. Because this interval is not in the asymptotic $T\rightarrow0$ limit, however, the fitted power law should not by itself be interpreted as evidence for a specific gap symmetry or microscopic penetration-depth mechanism.

\section{Discussion and Conclusions}
\begin{figure*}
	\centering
 	\includegraphics[width=6.2in]{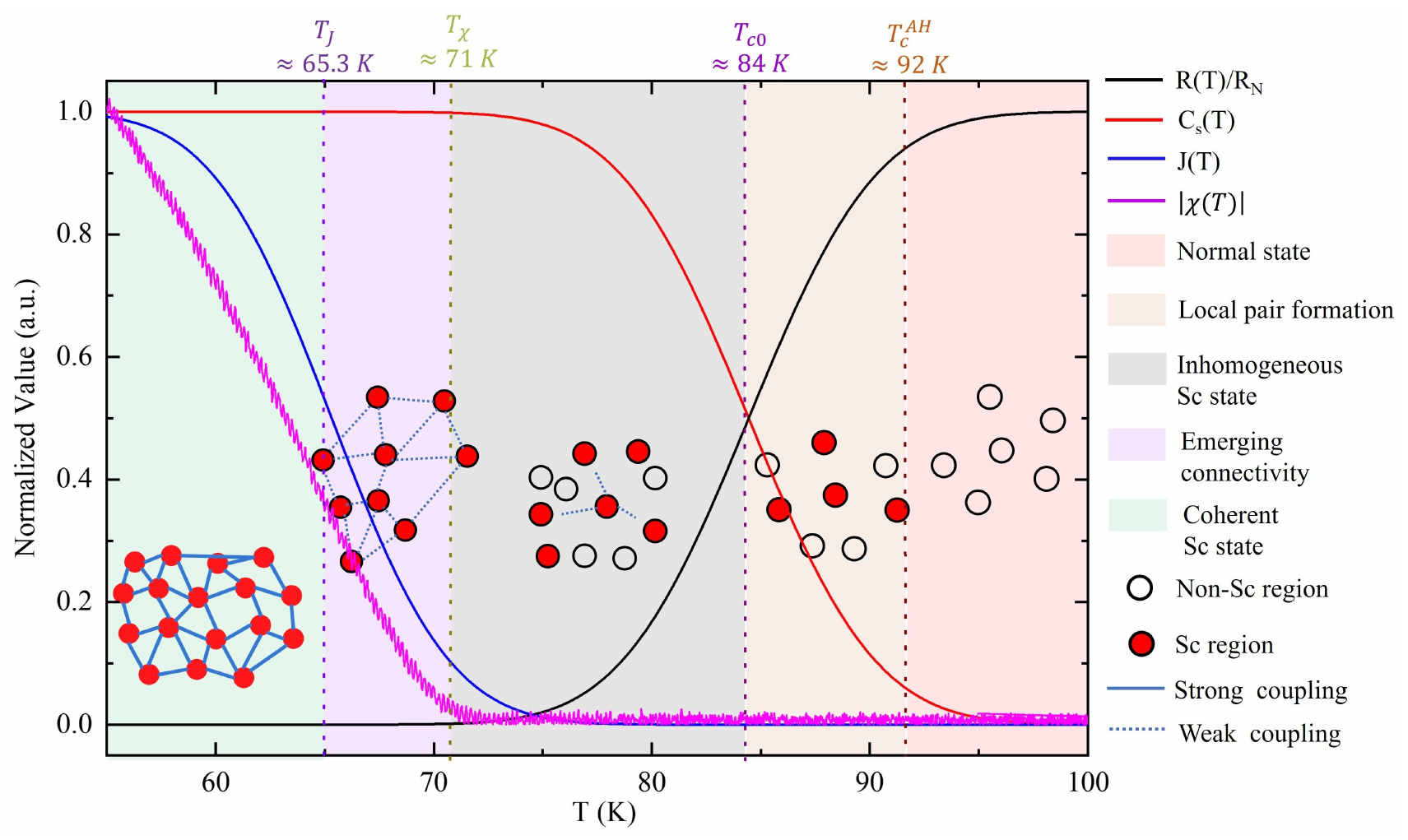}
	\caption{Schematic summary of the superconducting (Sc) evolution constructed from the combined transport and susceptibility analyses. The normalized resistance, superconducting fraction $c_s(T)$, connectivity factor $J(T)$, and diamagnetic response are shown schematically to illustrate the progressive development of superconductivity upon cooling. The characteristic temperatures $T_c^{AH}$, $T_{c0}$, $T_{\chi}$, and $T_J$ delineate distinct regimes associated with local Cooper-pair formation, growth of the superconducting fraction, emergence of diamagnetic screening, and establishment of Josephson-mediated phase coherence, respectively. The lower panels schematically illustrate the evolution of the superconducting network, where red and white circles represent superconducting and normal regions, respectively. Solid and dashed lines denote strong and weak Josephson coupling between neighbouring superconducting regions.}
	\label{fig:fig5}
\end{figure*}

Although the Gaussian $T_c$-distribution and Ambegaokar--Halperin (AH) models both reproduce the measured $R(T)$ transition, they describe complementary aspects of the superconducting state. The Gaussian model provides a statistical description of spatial variations in the local transition temperature and yields the superconducting fraction $C_s(T)$, whereas the AH model describes dissipation associated with thermally activated phase dynamics in an effective Josephson-coupled network. The coupling energy obtained from the AH analysis therefore characterizes an effective phase-stiffness scale rather than the intrinsic pairing strength. These two descriptions are physically connected because regions with different local superconducting properties can also exhibit different coupling strengths. As $C_s(T)$ increases upon cooling, the connectivity between superconducting regions develops and thermally activated phase fluctuations are progressively suppressed.

\par The evolution inferred from the combined transport and susceptibility analyses is summarized schematically in Fig.~\ref{fig:fig5}. The AH analysis gives an effective onset scale $T_c^{\mathrm{AH}}\approx92$~K, while the Gaussian analysis yields a mean local transition temperature $T_{c0}\approx84.4$~K. Detectable diamagnetic screening develops near $T_{\chi}\approx71$~K, and the modified Choy--Stoneham analysis gives a connectivity midpoint $T_J\approx65.3$~K. The hierarchy $T_c^{\mathrm{AH}}>T_{c0}>T_{\chi}>T_J$ indicates that the superconducting response develops progressively rather than at a single temperature. Local superconducting correlations initially appear, the superconducting fraction subsequently increases, and the gradual establishment of inter-region connectivity enables an increasingly strong macroscopic diamagnetic response.

\begin{table*}[t]
\centering
\caption{Comparison of superconducting transition parameters obtained from
the Gaussian-$T_c$ analysis of the present Bi-2212 film and representative
Bi-2212 systems reported in the literature.}
\label{tab:bi2212_rt_comparison}

\begin{threeparttable}
\footnotesize

\setlength{\tabcolsep}{4.0pt}
\renewcommand{\arraystretch}{1.16}

\begin{tabular}{@{}p{1.55cm}p{1.55cm}p{3.85cm}p{1.20cm}p{1.90cm}p{0.85cm}p{0.85cm}p{0.95cm}p{0.95cm}@{}}
\toprule

Reference
& Form
& Growth/preparation
& Thickness
& Substrate
& $T_{c0}$ (K)
& $\sigma_{T_c}$ (K)
& $\Delta T_{10-90}$ (K)
& $\sigma_{T_c}/T_{c0}$ (\%) \\

\midrule

Present work
& Thin film
& Sputtering
& $\sim$30 nm
& SrTiO$_3$
& 84.44 & 4.62 & 11.80 & 5.47 \\

Zhang \textit{et al.}\ \cite{Zhang2023}
& Thin film
& PLD
& NR
& SrTiO$_3$
& 93.80 & 4.77 & 12.20 & 5.08 \\

Zhang \textit{et al.}\ \cite{Zhang2023}
& Thin film
& PLD
& NR
& LaAlO$_3$
& 89.66 & 5.33 & 13.64 & 5.94 \\

Zhang \textit{et al.}\ \cite{Zhang2023}
& Thin film
& PLD
& NR
& DyScO$_3$
& 78.20 & 6.12 & 15.66 & 7.82 \\

De Vero \textit{et al.}\ \cite{DeVero}
& Thin film
& Infrared PLD
& 12~$\mu$m
& MgO (100)
& 39.50 & 1.74 & 4.45 & 4.40 \\

Mua \textit{et al.}\ \cite{Mua}
& Thin film
& KrF-excimer PLD
& 272 nm
& MgO (100)
& 69.30 & 4.57 & 11.69 & 6.59 \\

Liu \textit{et al.}\ \cite{Liu}
& Thin film
& Acetate-based sol--gel; 820\,$^{\circ}$C,
$p_{\mathrm{O}_2}=1.4$ kPa
& 222 nm
& LaAlO$_3$
& 86.47 & 3.57 & 9.13 & 4.12 \\

Simsek \textit{et al.}\ \cite{Simsek}
& Thick film
& Liquid-phase epitaxy
& $\sim$1~$\mu$m
& MgO (100)
& 82.48 & 2.17 & 5.55 & 2.63 \\

Jiang \textit{et al.}\ \cite{Jiang2014}
& Graphene-protected flake
& Mechanical exfoliation
& 3.5 u.c. $\approx$10.5 nm
& SiO$_2$/Si; graphene capped
& 87.27 & 4.22 & 10.80 & 4.83 \\

Vaid \textit{et al.}\ \cite{Vaid}
& Bulk crystal
& Self-flux pressure technique
& NR
& None; alumina crucible used
& 88.80 & 3.92 & 10.03 & 4.41 \\

Aytekin \textit{et al.}\ \cite{Aytekin2024}
& Polycrystalline bulk ceramic
& Solid-state reaction; Sn/Na-substituted Bi-2212, $x=0.25$
& Bulk pellet
& NR
& 80.16 & 4.30 & 11.00 & 5.36 \\

Aichner \textit{et al.}\ \cite{Aichner}
& Thin film
& PLD
& 46 nm
& LaAlO$_3$
& 93.17 & 5.99 & 15.33 & 6.43 \\

Yu \textit{et al.}\ \cite{Yu2019}
& Monolayer
& Mechanical exfoliation from bulk
Bi$_{1.9}$Sr$_{2.1}$CaCu$_2$O$_{8+\delta}$
& 1 layer $\approx$1.6 nm
& Si/285-nm SiO$_2$
& 90.40 & 4.74 & 12.13 & 5.24 \\

\bottomrule
\end{tabular}

\begin{tablenotes}[flushleft]
\footnotesize
\item NR: not reported; PLD: pulsed-laser deposition; u.c.: unit cell.
$T_{c0}$ and $\sigma_{T_c}$ were obtained using the same Gaussian-$T_c$
analysis. $\Delta T_{10-90}$ is the interval over which the resistance
decreases from 90\% to 10\% of the normal-state resistance.
\end{tablenotes}

\end{threeparttable}
\end{table*}

\begin{figure*}
	\centering
 	\includegraphics[width=6.2in]{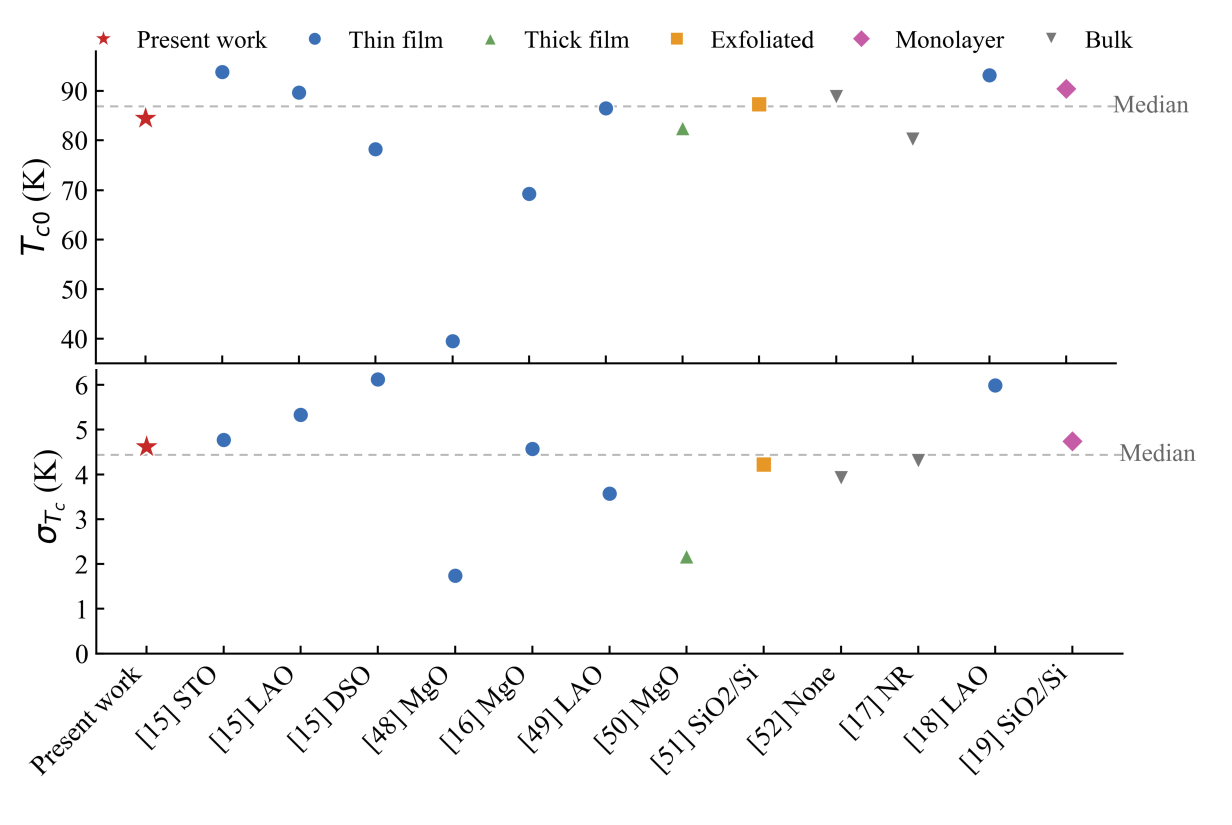}
	\caption{Comparison of the superconducting transition parameters obtained from the Gaussian-$T_c$ analysis of the present Bi-2212 film with representative Bi-2212 systems reported in the literature. (a) Mean superconducting transition temperature, $T_{c0}$, and (b) transition-width parameter, $\sigma_{T_c}$. The red star represents the present 30-nm sputtered Bi-2212 film, whereas the remaining symbols distinguish thin films, thick films, exfoliated flakes, monolayers, and bulk samples. The horizontal dashed line indicates the median value of the corresponding literature data. The reference numbers correspond to those listed in Table~\ref{tab:bi2212_rt_comparison}.}
	\label{fig:Tc_sigma_comparison}
\end{figure*}

To place the present transition in a broader context, published $R(T)$ curves for representative Bi-2212 samples were digitized and analyzed using the same Gaussian-$T_c$ model. The resulting $T_{c0}$, $\sigma_{T_c}$, and $\Delta T_{90\text{--}10}$ values are treated as effective measures of the position and width of each resistive transition. As summarized in Table~\ref{tab:bi2212_rt_comparison} and Fig.~\ref{fig:Tc_sigma_comparison}, the reported systems exhibit $T_{c0}$ values ranging from 39.50 to 93.80~K and $\sigma_{T_c}$ values ranging from 1.74 to 6.12~K. This variation reflects differences in sample geometry, growth method, thickness, substrate, oxygen stoichiometry, strain, disorder, and microstructure. For the present 30-nm-thick sputtered film, $T_{c0}=84.44$~K is close to the literature median of approximately 86.9~K, while $\sigma_{T_c}=4.62$~K is slightly larger than the corresponding median of approximately 4.44~K. Its normalized width, $\sigma_{T_c}/T_{c0}=5.47\%$, also lies within the range obtained for the other Bi-2212 systems. The present film therefore retains a relatively high transition temperature while exhibiting a degree of broadening comparable to that reported for Bi-2212 samples prepared by different routes.

\par The broader applicability of the transport--susceptibility framework was examined using measured data from a NbTiN film and previously reported transport and magnetic-response data for nanoporous NbN. As detailed in the Supplemental Material~\cite{Suppl}, these reference systems provide additional tests of the relationship between the growth of the superconducting fraction and the development of magnetic connectivity.

\par In conclusion, the broadened superconducting transition of the 30-nm-thick BSCCO film is consistently described by the combined effects of spatially varying local transition temperatures and thermally activated phase dynamics. Incorporating the transport-derived superconducting fraction into the modified Choy--Stoneham model, together with the connectivity factor $J(T)$, quantitatively relates the resistive and magnetic transitions. The analysis distinguishes the development of local superconductivity from the subsequent establishment of effective connectivity and macroscopic magnetic screening. This combined framework provides a practical approach for correlating transport and susceptibility responses in spatially inhomogeneous superconducting films.

\section*{ACKNOWLEDGMENTS}
S. P. J. acknowledges financial support from the Anusandhan National Research Foundation (ANRF), Government of India, through the National Post Doctoral Fellowship (File No.~PDF/2025/002400). A. G. acknowledges financial support from the Nano Mission, Department of Science and Technology, Government of India, under Grant No.~DST/NM/TUE/QM-10/2019.

\section*{APPENDIX A: CALCULATION OF $\Delta T_{90\text{-}10}$ AND DERIVATIVE WIDTH}
\begin{figure}
	\centering
 	\includegraphics[width=3.4in]{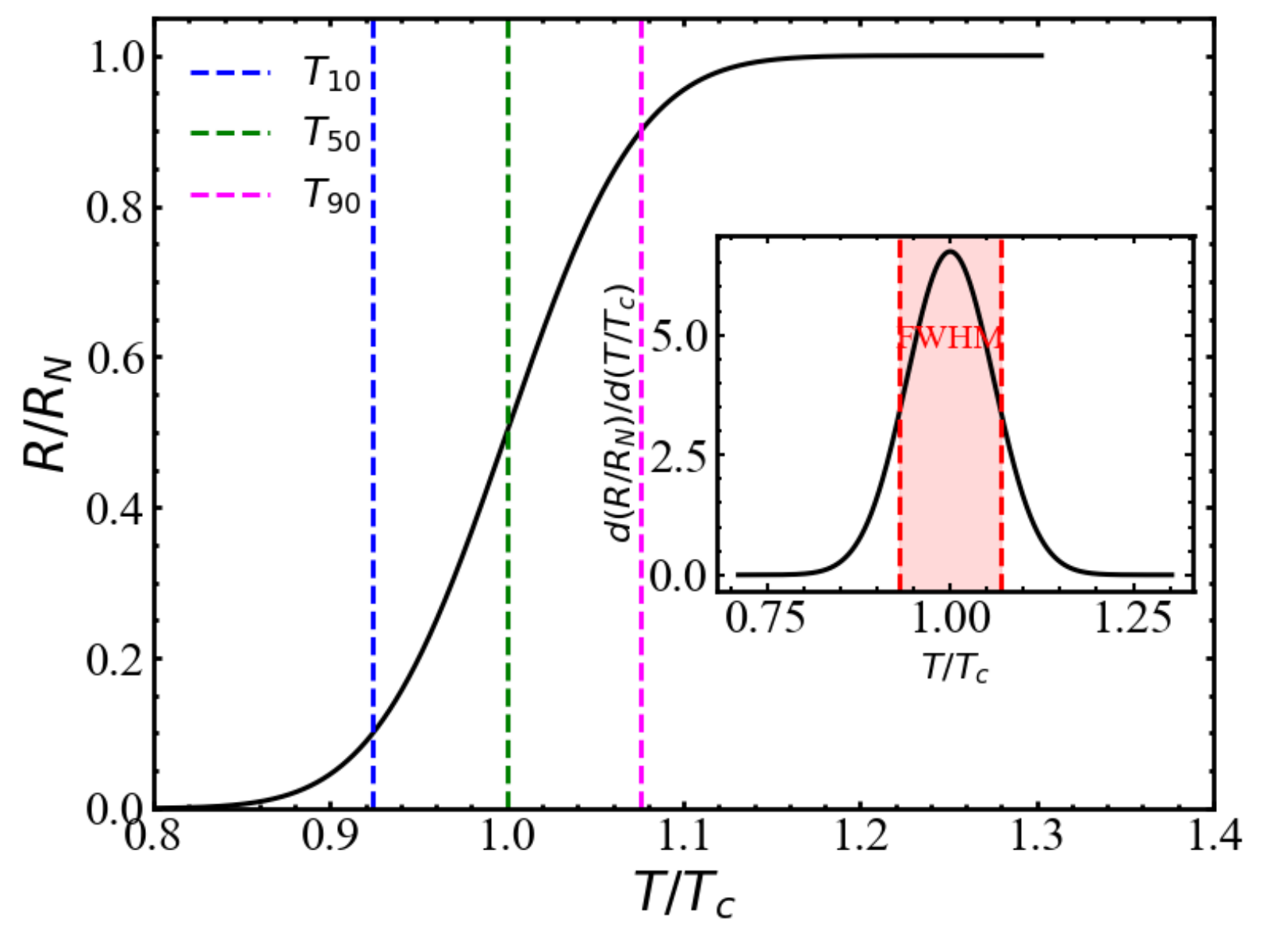}
	\caption{Normalized resistance $R/R_N$ as a function of reduced temperature $T/T_c$, showing the superconducting transition. The characteristic temperatures $T_{10}$, $T_{50}$, and $T_{90}$ are indicated by vertical dashed lines. Inset: normalized derivative $d(R/R_N)/d(T/T_c)$, where the shaded region denotes the FWHM of the transition.}
	\label{fig:fig6}
\end{figure}

The simulated superconducting transition is described using a Gaussian-broadened form of the resistance, $R(T)=\frac{R_N}{2}\left[1+\mathrm{erf}\left(\frac{T-T_{c0}}{\sqrt{2}\sigma}\right)\right]$, with the parameters defined earlier in the text. It is convenient to introduce the normalized resistance $r(T)=R(T)/R_N$, which gives $r(T)=\frac{1}{2}\left[1+\mathrm{erf}\left(\frac{T-T_{c0}}{\sqrt{2}\sigma}\right)\right]$. The characteristic temperatures $T_{10}$, $T_{50}$, and $T_{90}$ are defined from the conditions $r(T_{10})=0.1$, $r(T_{50})=0.5$, and $r(T_{90})=0.9$. Solving these relations gives $\mathrm{erf}\left(\frac{T-T_{c0}}{\sqrt{2}\sigma}\right)=2r-1$, which leads to $T = T_{c0} + \sqrt{2}\sigma\,\mathrm{erf}^{-1}(2r-1)$. Substituting $r=0.1$, $0.5$, and $0.9$ yields $T_{10}=T_{c0}+\sqrt{2}\sigma\,\mathrm{erf}^{-1}(-0.8)$, $T_{50}=T_{c0}$, and $T_{90}=T_{c0}+\sqrt{2}\sigma\,\mathrm{erf}^{-1}(0.8)$. Since $\mathrm{erf}^{-1}(0.8)=0.906$ \cite{error function}, the transition width becomes $\Delta T = T_{90}-T_{10} = 2\sqrt{2}\sigma\,\mathrm{erf}^{-1}(0.8) \approx 2.56\sigma$. For the value $\sigma=5$ K used in the simulation, this gives $\Delta T \approx 12.8$ K. Fig. \ref{fig:fig6} shows the simulated $R(T)$ curve with the characteristic temperatures $T_{10}$, $T_{50}$, and $T_{90}$ marked.
\par Further insight into the transition width can be obtained from the derivative of the resistance. Differentiating the model expression gives $\frac{dR}{dT}=\frac{R_N}{\sqrt{2\pi}\sigma}\exp\left[-\frac{(T-T_{c0})^2}{2\sigma^2}\right]$, which is a Gaussian centered at $T_{c0}$. The maximum occurs at $T=T_{c0}$. The full width at half maximum (FWHM) is obtained by solving $\frac{dR}{dT}=\frac{1}{2}\left(\frac{dR}{dT}\right)_{\mathrm{max}}$, which leads to $\exp\left[-\frac{(T-T_{c0})^2}{2\sigma^2}\right]=1/2$. Taking the logarithm gives $(T-T_{c0})^2=2\sigma^2\ln2$, so the half-maximum positions occur at $T=T_{c0}\pm\sigma\sqrt{2\ln2}$. The resulting width is $\mathrm{FWHM}=2\sigma\sqrt{2\ln2}\approx2.355\sigma$. For $\sigma=5$ K this yields $\mathrm{FWHM}\approx11.8$ K. Inset of Fig. \ref{fig:fig4} shows the derivative $dR/dT$ of the simulated transition, with the FWHM positions indicated.

\end{document}


\title{Supplementary Material on 
“Unified Transport and Susceptibility Analysis of a Thin BSCCO Film: From Local Pairing to Global Phase Coherence"}
\author{Santu Prasad Jana}
\affiliation{Centre for Nanoscience and Engineering, Indian Institute of Science, Bengaluru, Karnataka 560012, India}
\author{Bismaya Ranjan Nayak}
\affiliation{Department of Physics, Indian Institute of Science, Bengaluru, Karnataka 560012, India}
\author{Sohini Guin}
\affiliation{Centre for Nanoscience and Engineering, Indian Institute of Science, Bengaluru, Karnataka 560012, India}
\author{Akshay Naik}
\affiliation{Centre for Nanoscience and Engineering, Indian Institute of Science, Bengaluru, Karnataka 560012, India}
\author{Dhavala Suri}
\affiliation{Centre for Nanoscience and Engineering, Indian Institute of Science, Bengaluru, Karnataka 560012, India}
\author{Arindam Ghosh}
\affiliation{Department of Physics, Indian Institute of Science, Bengaluru, Karnataka 560012, India}
\date{\today}

\maketitle

To examine the applicability of the transport--susceptibility framework beyond the BSCCO film, the same analysis was applied to two reference superconducting systems. First, transport and susceptibility measurements were performed on a NbTiN film to verify the response of the two-coil mutual-inductance setup in a comparatively sharp superconducting transition. Second, previously reported transport and magnetic-response data for a nanoporous NbN film were reanalyzed \cite{Kumar2013}. In the latter system, the formation of superconducting islands and their subsequent collective coupling were identified independently from its nanoporous morphology and two-step transition. These systems provide useful reference cases for comparing the characteristic temperature scales obtained for BSCCO.

\subsection{NbTiN reference measurement}

\begin{figure*}
    \centering
    \includegraphics[width=6.6in]{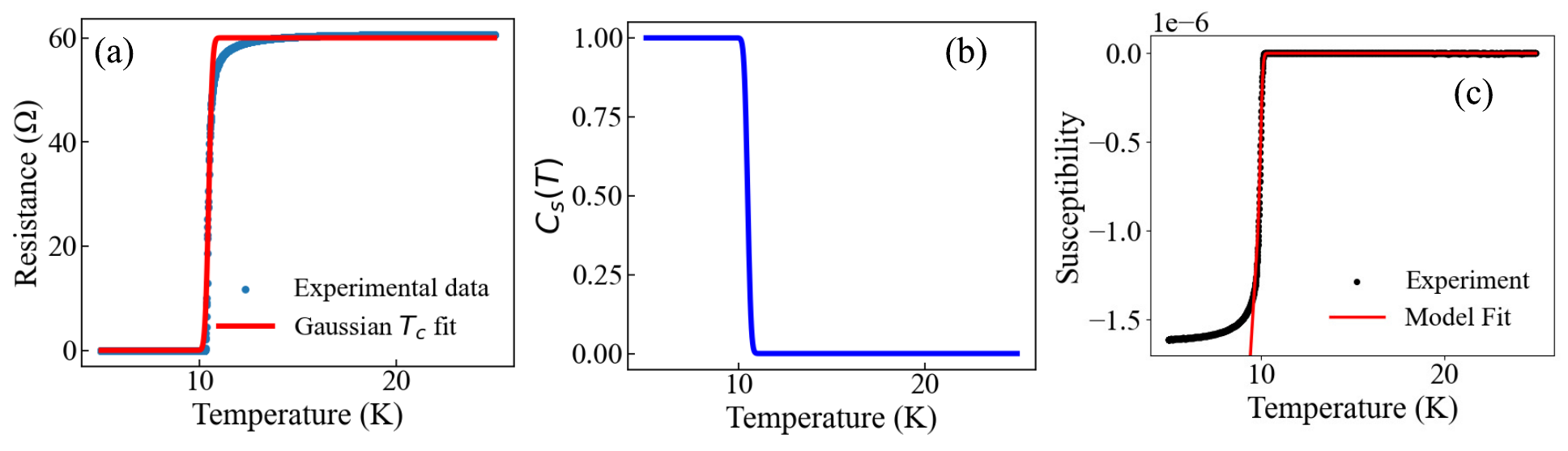}
    \caption{Transport and susceptibility characterization of the 25-nm-thick NbTiN reference film. (a) Temperature-dependent resistance and Gaussian $T_c$-distribution fit. (b) Superconducting fraction $C_s(T)$ obtained from the transport analysis. (c) Background-subtracted susceptibility and fit obtained using the connectivity-modified Choy--Stoneham model.}
    \label{fig:NbTiN}
\end{figure*}

Temperature-dependent resistance and susceptibility measurements were performed on a 25-nm-thick sputter-grown NbTiN reference film. As shown in Fig.~\ref{fig:NbTiN}(a), the film exhibits a sharp resistive superconducting transition. The Gaussian $T_c$-distribution analysis gives $T_{c0}=10.47$~K, $\sigma=0.15$~K, and $R_N=59.98~\Omega$. The corresponding superconducting fraction $C_s(T)$ is shown in Fig.~\ref{fig:NbTiN}(b). The small value of $\sigma$ indicates a narrow effective distribution of local transition temperatures compared with that obtained for the BSCCO film.

The background-subtracted susceptibility is shown in Fig.~\ref{fig:NbTiN}(c) together with the fit obtained using the connectivity-modified Choy--Stoneham model. The fit yields $T_J=10.04$~K, $\Delta_J=0.14$~K, and $x_0=a/\lambda_0=0.04678$. Taking $\lambda_0\approx400$~nm as a representative penetration-depth scale for NbTiN thin films \cite{Lee2024} gives an effective screening length $a\approx18.7$~nm. Because only the ratio $a/\lambda_0$ is constrained by the fit, this value should not be interpreted as a direct measurement of the physical grain or domain size.

The difference between the transport-derived mean transition temperature and the connectivity midpoint is $T_{c0}-T_J=0.43$~K. Thus, the development of the superconducting fraction and magnetic connectivity occurs over closely spaced temperature intervals in NbTiN. The measurement therefore verifies that the experimental setup and analysis recover nearly coincident transport and magnetic transitions in a superconductor with a comparatively sharp transition.

\subsection{Application to a nanoporous NbN film}

As a further test, the framework was applied to previously reported data for a 25--30-nm-thick nanoporous NbN film grown on an anodized alumina membrane \cite{Kumar2013}. The membrane had a pore diameter of approximately 25~nm and a pore-to-pore separation of 53~nm. The original study described the film as a network of weakly coupled superconducting regions produced by spatial variations in film thickness. Its upper resistive transition was associated with the formation of superconductivity within individual regions, whereas the lower transition was attributed to the development of collective phase coherence.

\begin{figure*}
    \centering
    \includegraphics[width=6.6in]{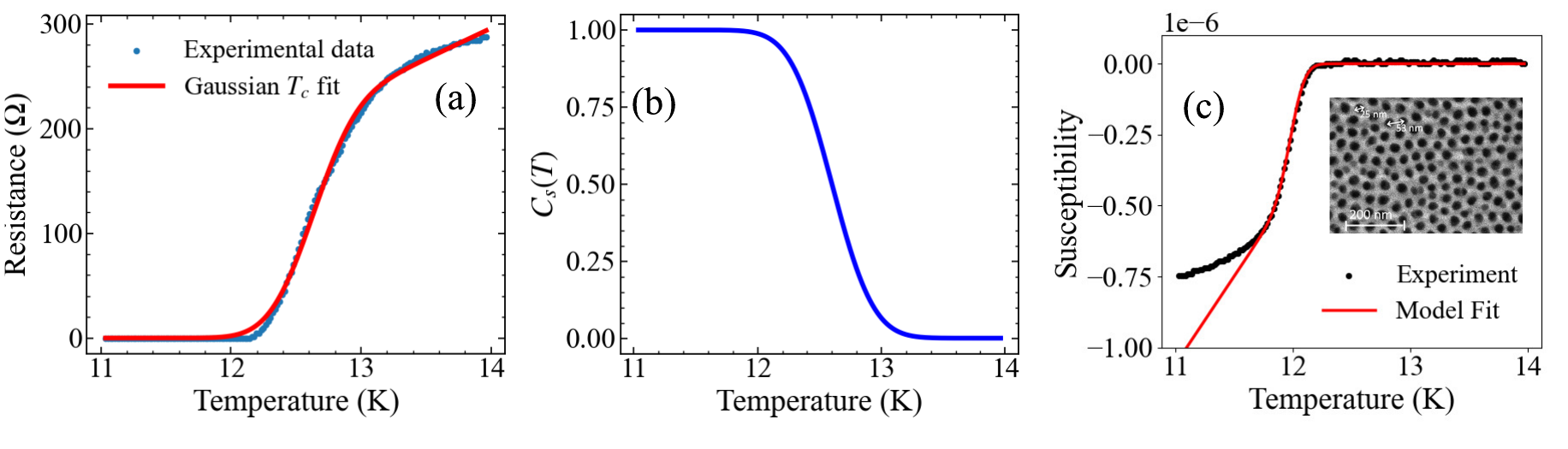}
    \caption{Application of the transport--susceptibility analysis to the nanoporous NbN film reported in Ref.~\protect\cite{Kumar2013}. (a) Digitized resistance data and Gaussian $T_c$-distribution fit. (b) Superconducting fraction $C_s(T)$ obtained from the transport analysis. (c) Reported magnetic response and fit obtained using the connectivity-modified Choy--Stoneham model. The inset shows an SEM image of the nanoporous film.}
    \label{fig:NbN}
\end{figure*}

Analysis of the digitized $R(T)$ data in Fig.~\ref{fig:NbN}(a) gives $T_{c0}=12.71$~K, $\sigma=0.34$~K, and $R_N=275.58~\Omega$. The corresponding $C_s(T)$, shown in Fig.~\ref{fig:NbN}(b), develops around the upper resistive transition. Applying the connectivity-modified susceptibility model to the magnetic-response data gives $T_J=12.00$~K, $\Delta_J=0.14$~K, and $x_0=a/\lambda_0=0.03143$. Taking $\lambda_0\approx400$~nm, consistent with the penetration-depth scale used for NbN in Ref.~\cite{Kumar2013}, gives an effective screening length $a\approx12.6$~nm. As in the BSCCO and NbTiN analyses, this value represents an effective model length rather than a direct determination of the superconducting-region size.

The connectivity midpoint lies below the mean resistive-transition temperature by $T_{c0}-T_J=0.71$~K. The analysis therefore recovers the ordering expected from the original interpretation: superconducting regions develop before their collective coupling becomes sufficiently strong to produce the principal magnetic response. Because the network morphology of this system is structurally evident, its behavior provides a useful external consistency check for the interpretation applied to the BSCCO film. It does not, however, establish that the microscopic weak-link structures are identical in the two materials.

\subsection{Comparison of characteristic temperature scales}

The parameters obtained for BSCCO, nanoporous NbN, and NbTiN are summarized in Table~\ref{tab:comparison}. The difference $T_{c0}-T_J$ measures the separation between the mean resistive-transition temperature and the connectivity midpoint. The quantities $\sigma$ and $\Delta_J$ characterize the breadth of the corresponding crossovers, although their numerical values are not identical statistical measures because they enter differently defined error-function expressions.

\begin{table}[t]
\caption{Characteristic parameters obtained from the Gaussian $T_c$-distribution and connectivity analyses for BSCCO, nanoporous NbN, and NbTiN. All quantities are given in kelvin.}
\label{tab:comparison}
\centering
\begin{ruledtabular}
\begin{tabular}{lccccc}
Material &
$T_{c0}$ &
$T_J$ &
$T_{c0}-T_J$ &
$\sigma$ &
$\Delta_J$ \\
BSCCO & 84.44 & 65.32 & 19.12 & 4.62 & 6.08 \\
NbN   & 12.71 & 12.00 & 0.71  & 0.34 & 0.14 \\
NbTiN & 10.47 & 10.04 & 0.43  & 0.15 & 0.14 \\
\end{tabular}
\end{ruledtabular}
\end{table}

NbTiN exhibits a narrow Gaussian distribution, $\sigma=0.15$~K, and a narrow connectivity crossover, $\Delta_J=0.14$~K, with $T_{c0}$ and $T_J$ separated by only 0.43~K. The nanoporous NbN film also shows a relatively small separation of 0.71~K. These separations correspond to approximately 4.1\% and 5.6\% of their respective $T_{c0}$ values.

In contrast, the BSCCO film exhibits larger transition-width parameters and a separation $T_{c0}-T_J=19.12$~K, corresponding to approximately 22.6\% of $T_{c0}$. The framework therefore distinguishes the closely spaced transport and connectivity scales of NbTiN and nanoporous NbN from the widely separated scales found in BSCCO. These reference analyses support the internal consistency and broader applicability of the framework while strengthening the interpretation that spatial inhomogeneity and delayed inter-region connectivity both contribute to the broad transport and magnetic transitions of the BSCCO film.